# Manipulating the Polar mismatch at LaNiO$_3$/SrTiO$_3$ (111) Interface

M. Saghayezhian[1], Zhen Wang[1,2], Hangwen Guo[1], Yimei Zhu[2], E.W. Plummer[1], Jiandi Zhang[1]

[1] Department of Physics and Astronomy, Louisiana State University, Baton Rouge, Louisiana 70803, USA.

[2] Condensed Matter Physics and Materials Science Department, Brookhaven National Laboratory, Upton, New York 11973, United States

Heteroepitaxial growth of transition-metal oxide films on the open (111) surface of SrTiO$_3$ results in significant restructuring due to the polar mismatch. Monitoring the structural and composition on an atomic scale of LaNiO$_3$/SrTiO$_3$ (111) interface as a function of processing conditions has enabled the avoidance of the expected polar catastrophe. Using atomically resolved transmission electron microscopy and spectroscopy as well as Low energy electron diffraction, the structure of the thin film, from interface to the surface, has been studied. In this paper, we show that the proper processing can lead to a structure that is ordered, coherent with the substrate without intermediate structural phase. Angle-resolved X-ray photoemission spectroscopy shows that the oxygen content of thin films increases with the film thickness, indicating that the polar mismatch is avoided by the presence of oxygen vacancies.

Transition metal oxide heterostructures exhibit variety of remarkable interfacial properties due to the lattice mismatch, orbital character, charge transfer, polar mismatch, or broken symmetry [1]. For example, the interface of LaAlO$_3$/SrTiO$_3$ (001) has shown that a 2-dimensional electron gas (2DEG), coexists with superconductivity and ferromagnetism [2-4]. These unusual interfacial phenomena have ignited tremendous effort aimed at engineering or controlling interface properties [5,6]. An important aspect of the search for and control of interfacial properties is the orientation of the substrate [7-9]. A prototype example is the LaNiO$_3$/LaMnO$_3$ superlattice in highly polar [111] direction, exhibiting an unusual coupling at the interface, which displays exchange bias between ferromagnetic LaMnO$_3$ and paramagnetic LaNiO$_3$ [10,11]. It has been predicted [12], though yet to be verified [13], that superlattices of LaNiO$_3$/LaAlO$_3$ (111) and LaNiO$_3$/SrTiO$_3$ (111) are host to topological interface states that show transition to Mott state. The ability to create the sharp interface in these systems opens up the possibility of controlling parameters such as interfacial correlations and coupling as well as tuning of crystal field using strain and interface directionality to manipulate intriguing properties [14-16].

Figure 1 illustrates the dramatic differences of LaNiO$_3$/SrTiO$_3$ interfaces depending upon the orientation of the STO substrate. In bulk, LaNiO$_3$ is a paramagnetic metal [17], where the nominal oxidation state of Ni is +3, with a low spin 3d$^7$ electronic configuration. As shown in Fig. 1(a) in the [111] direction, the stacking of Ni$^{3+}$ and (LaO$_3$)$^{3-}$ have in-plane uncompensated charge of 3+ and 3-, which makes [111] direction highly polar. The SrTiO$_3$ (111) substrate is formed by stacking of Ti$^{4+}$ and (SrO$_3$)$^{4-}$, which exhibits sequential repetition of in-plane net charge of 4+ and 4-, making SrTiO$_3$ even more polar than LaNiO$_3$. The charge imbalance at the interface, between Ti$^{4+}$ and (LaO$_3$)$^{3-}$ creates a discontinuity in the electric potential, hence a *polar discontinuity,* which results in divergence of the interface free energy. In [001] direction, as shown in Fig. 1(b), the substrate does not have uncompensated charge, therefore it is not polar. The (001) interface is considered as weakly polar due to the uncompensated charge on Nickelate side. The interplanar spacing for (111) is about 1.7 times smaller than for [001] direction making the interface region more susceptible to intermixing.

To avoid the divergence of energy at the interface, the interface either severely intermixes or the thin film grows in another phase, which in the case of LaNiO$_3$ is La$_2$Ni$_2$O$_5$ [18]. There have been several attempts to address the polar discontinuity issue [18-23], but producing a single-phase thin film with sharp interface still remains a challenge. In most cases, the first few layers grow in different, unwanted phases to prevent the divergence of surface free energy, usually referred to as polar catastrophe. There are four possible avenues to avoid polar catastrophe: adsorption (contamination), electronic reconstruction, geometrical reconstruction and stoichiometric reconfiguration [24]. Among these, only electronic reconstruction can preserve the phase and structure of thin film at the interface, since the other three will cause interface roughening. A microscopic understanding of the interface dynamics during initial stages of growth is crucial. We show that with a proper processing procedure it is possible to avoid polar catastrophe and obtain a single phase thin film with uniform structure. We show that with a proper processing procedure for substrate and thin film, it is possible to avoid polar catastrophe and obtain a single phase thin film with uniform structure.

High quality thin films of $LaNiO_3/SrTiO_3$ (111) are grown using UHV pulsed laser deposition. The growth was monitored by high pressure reflection high energy electron diffraction (RHEED). The substrate was prepared using a special method described elsewhere [25]. Laser pulses of 180 mJ at repetition rate of 10 Hz were focused on stoichiometric $LaNiO_3$ target. During the growth, the substrate was at 625 °C and the 6% oxygen/ozone mixture with partial pressure of 10 mTorr. The thickness of the thin films was determined by RHEED oscillation, shown in Fig. 2(a). The streak like RHEED pattern in the end of growth shows a two dimensional (2D) thin film growth (inset Fig. 2(a)). Clear RHEED oscillations are a direct indication of crystalline thin film growth, but they do not provide information about the phase or structure of thin film in the growth direction, especially near the buried interface. The samples grown for *in situ* measurements such as, angle-resolved X-ray photoemission spectroscopy (ARXPS), low energy electron diffraction (LEED) and RHEED, were grown on 0.1% Nd-doped $SrTiO_3$ (111) to achieve conductivity. The samples used for *ex-situ* high resolution transmission electron microscopy (STEM) measurement were grown on both doped and non-doped SrTiO$_3$ (111).

Fig. 2(b) displays high-angle annular dark field (HAADF) STEM image of our LaNiO$_3$/SrTiO$_3$ (111) interface taken along [1$\bar{1}$0] direction, showing a sharp interface and extremely well ordered epitaxial film. The thin film is fully strained and no obvious interface roughening is observed. The substrate is Ti terminated and the thin film growth begins with the LaO$_3$ layer. Fig. 1(c) shows the elemental electron energy loss spectroscopy (EELS) mapping, providing the chemical composition of the interface. The line profiles of EELS mapping indicate that the interface intermixing is limited to the two unit cells, particularly at transition metal ion site (B-site). The intermixture between Ti and Ni based on the variation of Ti EELS intensity is about 50% and 20% in the first and second unit cell, respectively. However, our analysis from Ti EELS spectra (Fig. S1) shows there is a slight variation of chemical valence of Ti ions diffusing in the LaNiO$_3$ film. The valence state of Ti reduces from +4 in SrTiO$_3$ to ~+3.8 in the first and second unit cell of the film. Although the inter-diffusion Ti into LaNiO$_3$ is not large, the Ti plays an important role in compensating structural and polar mismatch at the interface. The role of Ti here is twofold. First the larger Ti ionic radius can alleviate the tensile stress at the interface. Second, the partially occupied d-orbital of Ti at LaNiO$_3$ side will help screen the uncompensated charges at the interface. The two-unit cell inter-diffusion in [111] direction is about 0.44 nm which translates to about 1.2 LaNiO$_3$ unit cell in [001] direction. Considering atomic packing factor in [111] direction, this means that the LaNiO$_3$ (111) thin film has a sharp interface. The Ni ELS spectra was not recorded because the cross section of Ni 2p core level is very low and requires an intense electron beam. Increasing the beam intensity damages the sample. We have investigated the Ni oxidation state using ARXPS in the following.

In order to understand the evolution of the surface structure of thin films we studied the surface of 3 and 5 unit cell (uc) LaNiO$_3$ (111) using low energy electron diffraction (LEED), performed *in-situ* immediately after growth. Figures 3(a) and 3(b) show the LEED patterns for these two thin films. The sharp LEED spots confirm that the surface is well ordered. Both images exhibit three-fold symmetry, following the symmetry of the substrate and the symmetry expected for the epitaxial film (Fig.S2). The desired phase is LaNiO$_3$, but a previous study observed La$_2$Ni$_2$O$_5$ [18] phase near the interface as an intermediate phase in the growth. The difference between LaNiO$_3$ and La$_2$Ni$_2$O$_5$ is the ordered oxygen vacancy rows, as shown in Fig. 3(c) and 3(d). The La$_2$Ni$_2$O$_5$ surface should result in a $2 \times 1$ reconstruction, where the expected LEED patterns for each phase is shown in the insets of Fig. 3(c) and 3(d). If the La$_2$Ni$_2$O$_5$ phase were present, the fractional spots would have been present at the positions of the red circles in Fig. 3 (a) and (b). The absence of fractional spots means that the film has the symmetry of bulk, i.e. not La$_2$Ni$_2$O$_5$. This observation indicates that polarity compensation does not drive the thin film into a new phase with reconstructed surface for our growth conditions.

Ordered rows of oxygen vacancy distinguish La$_2$Ni$_2$O$_5$ from LaNiO$_3$. Since the Oxygen is a light element, we performed annular bright field (ABF) STEM imaging, which is sensitive to light elements. Figure 4(a) is the ABF-STEM image of the LaNiO$_3$/ SrTiO$_3$ interface for a 16uc LaNiO$_3$ film. The Fast Fourier transform (FFT) of the ABF STEM image of the LaNiO$_3$ film is shown in Fig. 4(b). This diffraction pattern can be compared to what would be expected for the two different phases, La$_2$Ni$_2$O$_5$ or LaNiO$_3$. Fig. 4 (c) and (d) marble models of the two different structures projected along [$1\bar{1}0$]. The insets in the Fig. 4 (c) and (d) show simulated electron diffraction pattern of the ideal structure. Presence of ordered rows of Oxygen vacancies for the La$_2$Ni$_2$O$_5$ structure (Fig. 4(d)) results in the presence of fractional order spots. The red circles in Fig. 4 indicate the position where the fractional order spots should appear, but the spots are missing. The advantage of this method is that one can take Fourier transform of different areas of the thin film to see if there are patches of $La_2Ni_2O_5$ co-existing with $LaNiO_3$ phase, which was never observed.

We utilized XPS to study the oxidation states of Ni for four film thicknesses (5, 7, 9 and 16 uc). Figure 5a-d displays the data for the Ni 3p core level spectra at normal emission for different thicknesses. Normal emission was chosen to maximize the depth sensitivity of XPS. The spectra were fitted to four Gaussian-Lorentzian peaks, which represent two oxidation states of Ni (3+ and 2+) and two spin-orbit components of each oxidation state ($\frac{1}{2}$ and $\frac{3}{2}$) [26]. To minimize the number of free parameters used in the fitting and increase the reliability of results, the branching ratio (1:2), spin-orbit splitting energy (2 eV) [27], and FWHM of the peaks were held constant. Fig. 5(e) illustrates that with increasing thickness, the ratio of $Ni^{3+}/Ni^{2+}$ peak intensity increases (blue curve). Using this ratio, we can calculate the nominal amount of oxygen vacancies by fixing the stoichiometry according to formula $LaNiO_x$. The resulting $x$ equals are 2.55, 2.61, 2.63 and 2.65 for 5, 7, 9 and 16 uc thick films, respectively. The calculated oxygen content of thin films based on our XPS results approaches the oxygen content of $La_2Ni_2O_5$ ($x = 2.5$) with decreasing film thickness, but there is no indication of the existence of $La_2Ni_2O_5$ phase in the HAADF-STEM results. The binding energy of the Ni 3p core level, shown in Fig. 5(e) appears to exhibit a sudden shift to lower energy for films thicker than 7 uc. The same behavior was observed in the O 1s and La 4d core levels (Fig. S3). This means the core hole screening increases for thicknesses above 7 uc. The enhanced core hole screening is an indication of enhanced metallicity [28]. This is consistent with the fact that with increasing the thickness, the amount of oxygen vacancies decrease which restores the metallicity of $LaNiO_3$, and agrees with the previous work where it was shown that with increasing thickness the metallicity of the thin film increases [18].

In order to resolve the puzzle of the thickness-dependent oxygen content vs. no structural change, it is important to know the distribution of the oxygen vacancies. Are they uniform throughout the thin film or are they concentrated near the interface? We have performed large angle XPS to enhance the surface sensitivity. Fig. 5(f) shows that there is no measurable difference between binding energy of O 1s, Ni 3p and La 4d core levels at normal emission compared to $\theta = 75°$ emission angle. If the chemical environment of O, Ni and La in the film differed from the region near the surface, then the initial state effect would cause a core level shift for these elements. The line shape of Ni 3p spectra for normal emission and $\theta = 75°$ are identical (Fig. S4). This result is consistent with elemental EELS analysis from HRTEM where no appreciable change was observed in the line shape and energy of EELS spectra of O K-edge.

In summary, ultra-thin films of LaNiO$_3$ have been grown epitaxially on SrTiO$_3$ in highly polar [111] direction. Structure and stoichiometry of the ultra-thin films has been systematically studied using a series of *in-situ* (RHEED, LEED, XPS) and *ex-situ* (STEM/EELS) techniques. There is no obvious interface roughening and cationic intermixture is limited to the first two unit cells, which shows that we achieved a coherent growth with a single-phase. Our results show that even in the presence strong polar discontinuity, it is possible to fabricate the desired digital superlattices.

*Acknowledgments*. This work is primarily supported by the US Department of Energy (DOE) under Grant No. DOE DE-SC0002136. The electronic microscopic work done at Brookhaven National Laboratory is sponsored by the US DOE Basic Energy Sciences, Materials Sciences and Engineering Division under Contract DE-AC02-98CH10886.

[1]     P. Zubko, S. Gariglio, M. Gabay, P. Ghosez, and J.-M. Triscone, Annu. Rev. Condens. Matter Phys. **2**, 141 (2011).
[2]     J. A. Bert, B. Kalisky, C. Bell, M. Kim, Y. Hikita, H. Y. Hwang, and K. A. Moler, Nature physics **7**, 767 (2011).
[3]     A. Ohtomo and H. Hwang, Nature **427**, 423 (2004).

[4]     N. Reyren *et al.*, Science **317**, 1196 (2007).
[5]     P. Zubko, S. Gariglio, M. Gabay, P. Ghosez, and J.-M. Triscone, Annual Review of Condensed Matter Physics **2**, 141 (2011).
[6]     H. Y. Hwang, Y. Iwasa, M. Kawasaki, B. Keimer, N. Nagaosa, and Y. Tokura, Nat Mater **11**, 103 (2012).
[7]     J. Chakhalian *et al.*, Physical Review Letters **107**, 116805 (2011).
[8]     J. H. Ngai, F. J. Walker, and C. H. Ahn, Annual Review of Materials Research **44**, 1 (2014).
[9]     M. Gibert *et al.*, Nat Commun **7** (2016).
[10]    M. Gibert, P. Zubko, R. Scherwitzl, J. Íñiguez, and J.-M. Triscone, Nature materials **11**, 195 (2012).
[11]    M. Gibert *et al.*, Nano Letters **15**, 7355 (2015).
[12]    D. Doennig, W. E. Pickett, and R. Pentcheva, Physical Review B **89**, 121110 (2014).
[13]    H. Wei, M. Grundmann, and M. Lorenz, Applied Physics Letters **109**, 082108 (2016).
[14]    D. Doennig, W. E. Pickett, and R. Pentcheva, Physical review letters **111**, 126804 (2013).
[15]    J. Chakhalian, A. Millis, and J. Rondinelli, Nature materials **11**, 92 (2012).
[16]    D. Xiao, W. Zhu, Y. Ran, N. Nagaosa, and S. Okamoto, Nat Commun **2**, 596 (2011).
[17]    G. Catalan, Phase Transitions **81**, 729 (2008).
[18]    S. Middey, P. Rivero, D. Meyers, M. Kareev, X. Liu, Y. Cao, J. Freeland, S. Barraza-Lopez, and J. Chakhalian, Scientific reports **4** (2014).
[19]    J. Blok, X. Wan, G. Koster, D. Blank, and G. Rijnders, Applied Physics Letters **99**, 151917 (2011).
[20]    I. Hallsteinsen *et al.*, Journal of Applied Physics **113**, 183512 (2013).
[21]    N. Nakagawa, H. Y. Hwang, and D. A. Muller, Nat Mater **5**, 204 (2006).
[22]    A. Savoia *et al.*, Physical Review B **80**, 075110 (2009).
[23]    S. Middey, J. Chakhalian, P. Mahadevan, J. W. Freeland, A. J. Millis, and D. D. Sarma, Annual Review of Materials Research **46**, 305 (2016).
[24]    N. Claudine, Journal of Physics: Condensed Matter **12**, R367 (2000).
[25]    M. Saghayezhian, L. Chen, G. Wang, H. Guo, E. W. Plummer, and J. Zhang, Physical Review B **93**, 125408 (2016).
[26]    M. P. Seah, Surface and Interface Analysis **2**, 222 (1980).
[27]    Q. Liang and B. Xiaofang, EPL (Europhysics Letters) **93**, 57002 (2011).
[28]    P. S. Bagus, E. S. Ilton, and C. J. Nelin, Surface Science Reports **68**, 273 (2013).

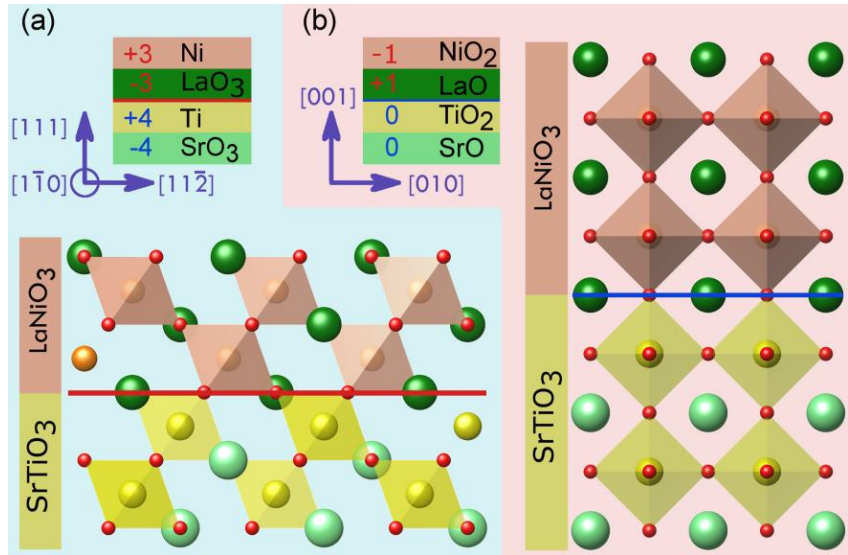

**Fig. 1** Stacking sequence of LaNiO$_3$/SrTiO$_3$ in (a) [111] and in (b) [001] direction. The structure of LaNiO$_3$/SrTiO$_3$ in [111] and [001] direction is shown, respectively. It is easily seen that packing factor of [111] direction is considerably larger than [001] direction.

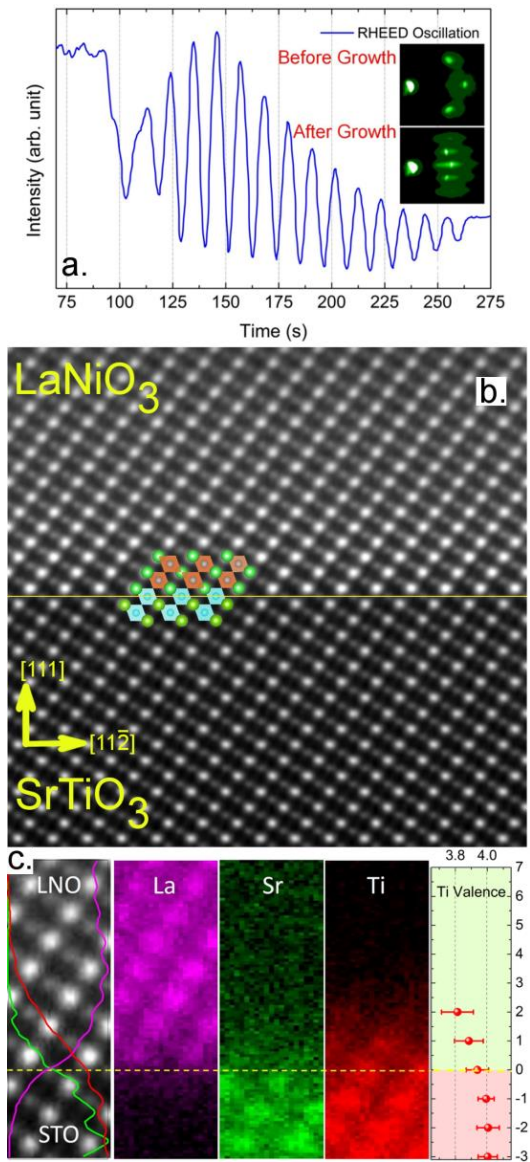

**Fig. 2** a) RHEED oscillations for LaNiO$_3$/SrTiO$_3$ (111) is presented for 15 u.c. The inset shows the RHEED pattern before and after growth. The streak-like pattern after growth is an indication of 2D growth mode. b) HAADF-STEM image of LaNiO$_3$/SrTiO$_3$ (111) along [1$\bar{1}$0] direction. The interface is marked by the yellow line and the ball model mapped on the image shows the schematic of LaNiO$_3$/SrTiO$_3$ (111). c) The EELS elemental mapping and line profiles for Ti, Sr and La. The change in the formal valence of Ti is shown across the interface.

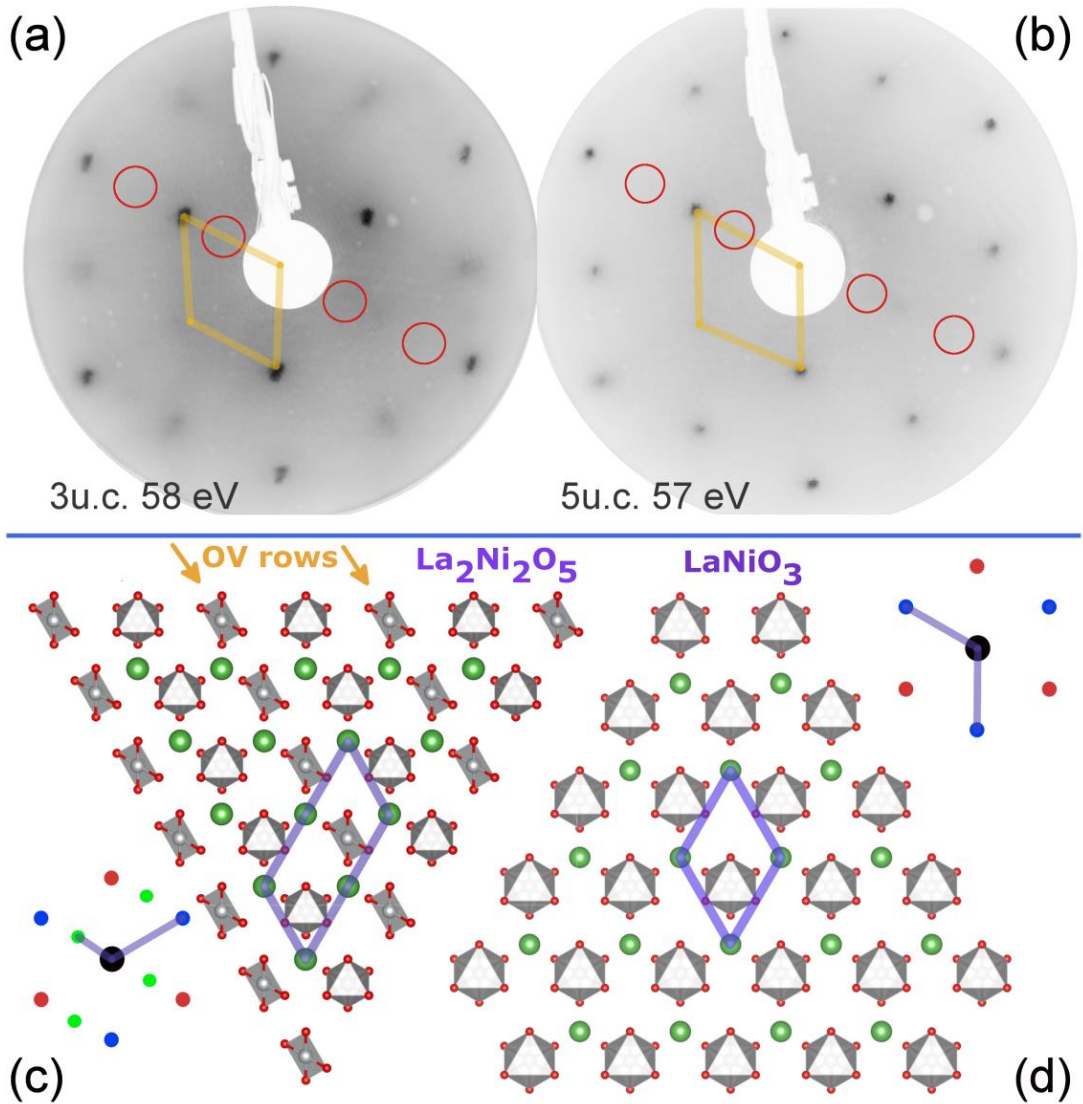

**Fig. 3** *a-b)* LEED pattern of 3 and 5 uc LaNiO3 (111) thin films. Empty red circles show the position of fractional spots which would be associated with a $La_2Ni_2O_5$ surface. c) Surface of $La_2Ni_2O_5$ (111). The rows of Oxygen vacancies are shown with yellow arrow. The simulated LEED pattern for this surface is shown. d) Surface of $LaNiO_3$ (111). The simulated LEED patterns are shown next to each structure. For simulated LEED patterns, green spots are fractional. Red and blue spots are integer, each color accounting for a domain.

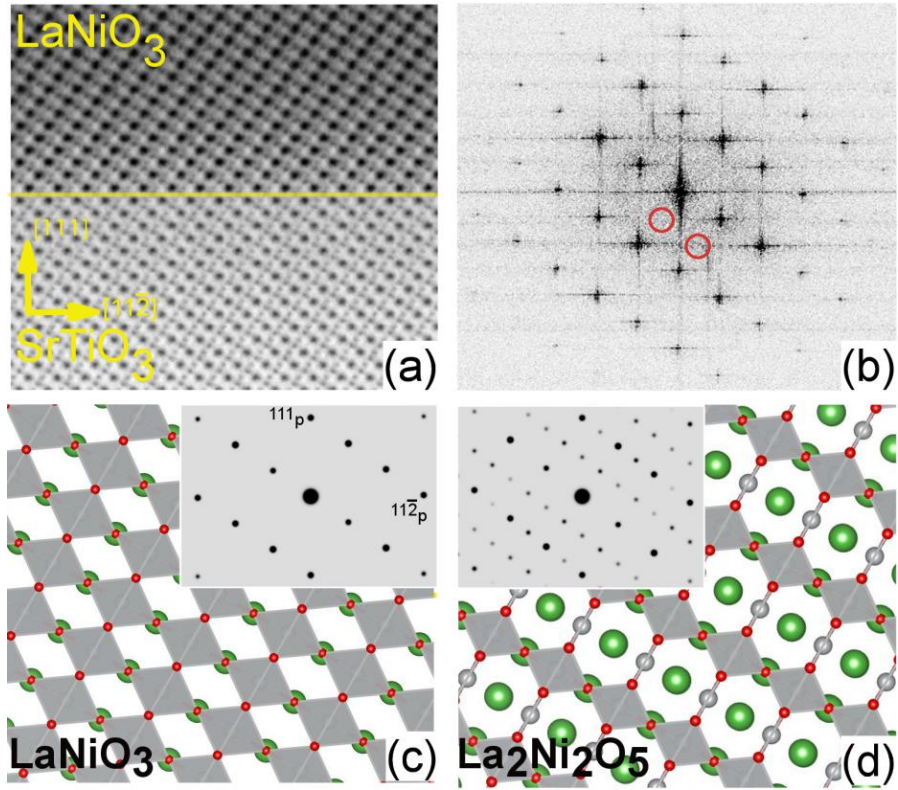

**Fig. 4** a) ABF-STEM image of LaNiO$_3$/SrTiO$_3$ (111) along [1$\bar{1}$0] direction. The interface is marked by the yellow line. b) Fast Fourier Transform (FFT) of the ABF-STEM image. Red circles indicate the position of fractional spots for La$_2$Ni$_2$O$_5$ phase. Absence of fractional spots in the FFT image indicates no ordered Oxygen vacancy. c-d) Schematic of LaNiO$_3$ and La$_2$Ni$_2$O$_5$ projected along [1$\bar{1}$0]. The simulated electron diffraction patterns are shown in the inset, respectively. The Fourier transform of La$_2$Ni$_2$O$_5$ shows fractional spots which are absent in LaNiO$_3$.

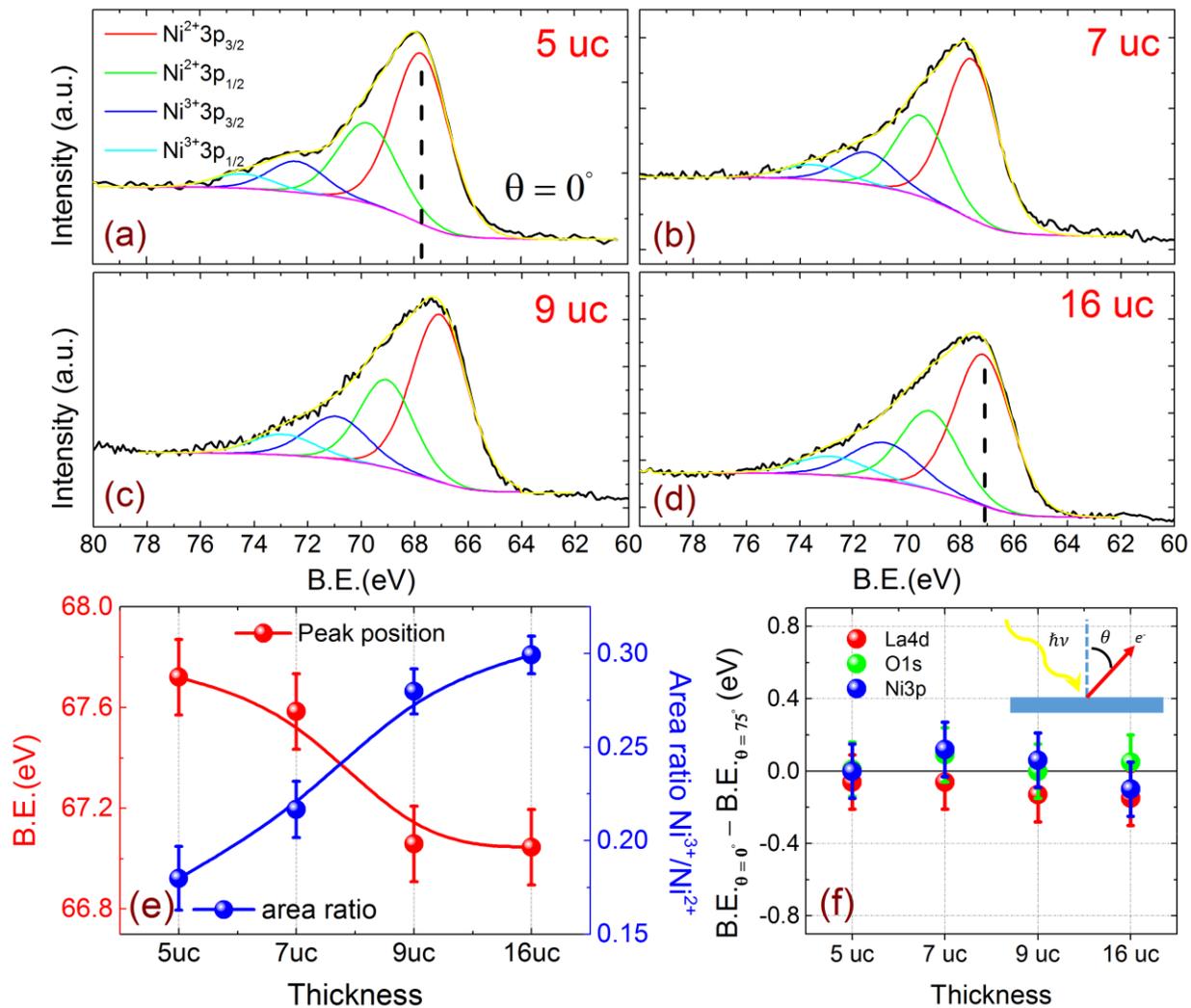

**Fig. 5** a-d) ARXPS spectra of Ni 3p for different thicknesses at normal emission. e) (*Left*) Change in binding energy of Ni 3p core level as a function of thickness and (*Right*) Change in area ratio of $Ni^{3+}/Ni^{2+}$ for Ni 3p core level. f) The difference in binding energy of La 4d, O 1s and Ni 3p in normal emission and $\theta = 75°$. The inset shows the schematic angle dependent XPS.

13